\documentstyle[12pt,aaspp4]{article}
\received{25 February 1997}
\accepted{23 April 1997}

\begin{document}

\title{Nebular Abundance Errors}
\author{Jason Alexander and Bruce Balick}
\affil{Astronomy Department, University of Washington,
    Seattle, WA 98195\\
Email: jalex@astro.washington.edu,balick@astro.washington.edu}

\begin{abstract}

  The errors inherent to the use of the standard ``ionization correction
  factor'' (``$i_{CF}$'') method of calculating nebular conditions and
  relative abundances of H, He, N, O, Ne, S, and Ar in emission line
  nebulae have been investigated under conditions typical for planetary
  nebulae.  The photoionization code {\sc Cloudy} was used to construct
  a series of model nebulae with properties spanning the range typical
  of PNe.  Its radial ``profiles'' of bright, frequently observed
  optical emission lines were then summed over a variety of
  ``apertures'' to generate sets of emission line measurements.  These
  resulting line ratios were processed using the $i_{CF}$ method to
  ``derive'' nebular conditions and abundances.  We find that for lines
  which are summed over the entire nebula the $i_{CF}$--derived
  abundances differ from the input abundances by $\leq 5\%$ for He and O
  up to $\leq 25\%$ or more for Ne, S, and Ar.  For resolved
  observations, however, the discrepancies are often much larger and are
  systematically variable with radius.  This effect is especially
  pronounced in low--ionization zones where nitrogen and oxygen are
  neutral or once--ionized such as in FLIERs, ansae and ionization
  fronts.  We argue that the reports of stellar--enriched N in the
  FLIERs of several PNe are probably specious.

\end{abstract}

\section{Introduction}

Chemical abundances of gaseous nebulae are among the most frequently
used astrophysical tools for measuring the rates of heavy element
formation, inferring the history of star formation, and gauging the
effects of internal stellar nucleosynthesis and mixing.  In this paper
we subject the most common method of measuring heavy element abundances,
the ``ionization correction factor'' (``$i_{CF}$'') method, to a simple
but rigorous examination.

The $i_{CF}$ method was initially developed by Torres--Peimbert \&
Peimbert (1979) as a method for converting measured nebular emission
line fluxes into chemical abundances.  The application of the method is
straightforward: Using standard diagnostic techniques, the electron
density, $n$, temperature, $T$, and the ionic abundances (relative to
H$^{+}$) of species with visible lines (e.g.  He$^{+}$, He$^{++}$,
N$^{+}$, O$^{+}$, O$^{++}$, Ne$^{++}$, S$^{+}$, S$^{++}$, Ar$^{++}$) are
derived from observed line ratios.  Total chemical abundances are then
derived by correcting the ionic abundances for unseen ionization stages
using $i_{CF}$s.

The many untestable assumptions used to formulate these correction
factors render them the weakest link in this methodology; however, the
resulting errors have never been stringently assessed.  Various
statistical studies of nebulae have been used to refine the various
correction factors of the $i_{CF}$ method, the latest and most
exhaustive study being that of Kingsburgh \& Barlow (1994) (hereafter,
KB).  However, these studies all rely on observations of nebulae whose
assumed chemical properties are based on $i_{CF}$ techniques -- a
somewhat circular process.  More recently, Henry {\it et~al}.\ (1996),
Kwitter \& Henry (1996), and Kwitter \& Henry (1997) have developed a
method in which $i_{CF}$ abundances are only used as a starting point
for {\sc Cloudy} models (see Ferland {\it et~al}.\ 1996 or Ferland {\it
  et~al}.\ 1994 for a description of {\sc Cloudy}).  Making certain
assumptions about the nebular structure and stellar properties, Henry
{\it et~al}.\ (1996) vary the input abundances to {\sc Cloudy} until its
predicted line ratios match their observed ones.  They find that
averaged over the nebula, $i_{CF}$ abundances differ by $\leq20\%$ from
those used in their most successful {\sc Cloudy} models.

If this were the entire story then this paper would not have been
submitted.  However, Balick {\it et~al}.\ (1994) observed several
planetary nebulae (PNe) with $1\farcs5$ resolution along a slit.  Using
the $i_{CF}$ method they derived radial increases in nitrogen abundances
as high as a factor of 7, whereas other elemental abundances remain
constant.  In addition, Guerrero {\it et~al}.\ (1995) have drawn similar
conclusions about nitrogen variations in the bipolar nebula M1-75 while
Corradi {\it et~al}.\ (1997) found radial increases for helium,
neon, sulfur, and argon in their long--slit studies of IC~4406.  Based
on such abundance variations with radius some authors have proposed the
recent ejection of high--velocity, element--enriched material by the
central star.  If real, these interpretations suggest some bizarre
stellar enrichment process for which no theoretical understanding
exists.

Alternately, the $i_{CF}$--derived abundance gradients could be
erroneous.  Investigating this possibility is the goal of the present
paper.  As we shall see, {\sc Cloudy} simulations show that there are
important regions, such as all low--ionization zones, in which the
$i_{CF}$ abundances of some elements can be highly discrepant (a factor
of three or more).  Other total abundances befall similar though less
extreme fates.

\section{Methodology}

We denote elements by standard abbreviations, H, He, C, etc.  Unless
implied otherwise in context, we shall use the following abbreviations
for specific emission lines: [N{\sc ii}] = [N{\sc ii}]$\lambda$6583\AA,
[O{\sc i}] = [O{\sc i}]$\lambda6300$\AA, [O{\sc ii}] = [O{\sc
  ii}]$\lambda3727$\AA = [O{\sc ii}]$\lambda\lambda(3726+3729)$\AA,
[O{\sc iii}] = [O{\sc iii}]$\lambda5007$\AA, [Ne{\sc iii}] = [Ne{\sc
  iii}]$\lambda3869$\AA, [Ar{\sc iii}] = [Ar{\sc iii}]$\lambda7136$\AA,
[Ar{\sc iv}] = [Ar{\sc iv}]$\lambda4740$\AA, [Ar{\sc v}] = [Ar{\sc
  v}]$\lambda7005$\AA, [S{\sc ii}] = [S{\sc
  ii}]$\lambda\lambda(6717+6731)$\AA, and [S{\sc iii}] = [S{\sc
  iii}]$\lambda6312$\AA.  Generally speaking, total abundances are
relative to H and ionic abundances are relative to H$^{+}$, so that
``Ne'' means the Ne/H abundance and ``O$^{++}$'' is equivalent to
O$^{++}$/H$^{+}$, etc.  ``Profile'' implies the radial distribution of
some quantity, e.g.\ an emission line profile.  The electron density,
$n({\rm S}^+)$, is determined using the 6717\AA\ and 6731\AA\ lines of
S$^+$ and the electron temperature is determined two ways; $T({\rm
  N}^+)$ using the [N{\sc ii}] 6583\AA, 6548\AA\ and 5755\AA\ lines and
$T({\rm O}^{++})$ using the [O{\sc iii}] 4959\AA, 5007\AA\ and 4363\AA\ 
lines using standard methods.

In the present study we used {\sc Cloudy} (version 90.02) in order to
conduct a controlled study of $i_{CF}$--derived abundances and their
sensitivity to changes in nebular conditions such as nebular geometry,
stellar temperature, luminosity and emergent spectrum, and abundances.
This procedure bypasses all observational problems (e.g.\ reddening
corrections and flux calibrations) and does not rely on a statistical or
empirical analysis of the observations of a heterogeneous sample of
observed objects.

Based on a set of ``{\it assumables}'', namely, the chemical abundances
$A(X)$ relative to H, along with a nebular density structure,
$n({\rm H},r)$, and a standard star model, $Q_*(T_*,L_*)$ or $Q_*(T_*,
g_*)$, {\sc Cloudy} was used to compute the profiles of the ionization
structure, electron density, electron temperature, and the emissivities
of bright optical lines.  {\sc Cloudy}'s emissivity profiles were summed
over different ``apertures'' to synthesize a set of ``{\it
  observables}'', i.e.\ emission lines fluxes in relatively bright
nebular lines.

The original profiles and these observables were normalized to H$\beta$
and then used to determine a set of ``{\it derivables}'' (denoted with a
prime) from ratios of observable lines -- $n'({\rm S}^+)$, $T'({\rm
  N}^+)$, $T'({\rm O}^{++})$, and ionic abundances $A'(X_{ion})$ --
using standard techniques.  The expressions for this process were
obtained by fits to output from the {\sc Ionic} and {\sc Temden} tasks
in the IRAF/STSDAS\footnote{Image Reduction and Analysis Facility is
  distributed by National Optical Astronomy Observatories, which is
  operated by the Association of Universities for Research in Astronomy,
  Inc. under cooperative agreement with the National Science
  Foundation.} program {\sc Nebular} (Shaw \& Dufour 1995).  These fits,
which are generally accurate to 2\% or better, are listed in Appendix A.
In all cases we used $T({\rm N}^+)$ and $T({\rm O}^{++})$ for computing
the abundances of ions emitted from low-- and high--ionization regions,
respectively.

The derived ionic abundances, $A'(X_{ion})$, were used as input to the
$i_{CF}$ method used by KB in order to derive the resultant total
abundances $A'(X)$.  To this end, the $A'(X_{ion})$ are multiplied by
certain $i_{CF}$ corrections, $i_{CF}(X)$, which are based on geometric
assumptions about the relative volumes occupied by the unseen ionization
stages.  Expressions for the $i_{CF}$ correction factors for various
atoms are also listed in the Appendix.  Finally, the ``discrepancy
ratios'', ${\cal A}(X) \equiv A'(X) / A(X)$ are computed.  (Note that
our ${\cal A}(X) \equiv 1/\xi(X)$ of Henry {\it et~al}.\ 1996).

As an example of this process, in many nebulae the dominant ionization
stages of nitrogen are N$^{+}$ and N$^{++}$.  Although N$^{+}$ emits
several bright optical lines (rendering N$^{+}$/H$^{+}$ readily
measurable), N$^{++}$ does not.  Hence we must estimate N by correcting
N$^{+}$ with an $i_{CF}$ where $i_{CF}($N$) =$N$ /$N$^{+}$, recalling of
course that N is not directly measurable.  In this case, the $i_{CF}$
method assumes that both nitrogen and oxygen have the same ionization
structure; i.e. the fractional ionizations $x($N$^{+}) = x($O$^{+})$ and
$x($N$^{++}) = x($O$^{++})$ at all radii.  Thus $i_{CF}$(N) = O/O$^{+}$,
where (O/O$^{+}$) can be derived from observables, as shown in Appendix
A.  Note that no explicit knowledge of the nebular physical conditions
or the excitation source is required beyond that needed to determine
ionic abundances.

Using this technique, an assessment of the global $i_{CF}$ nebular
abundances was made by integrating the line emissivity profiles computed
by {\sc Cloudy} over four regions or apertures: the entire model nebula
(truncated where $x($H$^{+}) \leq 30\%$); a ``high-ionization'' region
(in which $x($O$^{+}+$O$^{\rm o})<50\%$), a ``low-ionization'' region
(in which $x($O$^{+}) \geq 50\%$), and a ``neutral'' region (in which
$x($O$^{\rm o}) \geq 50\%$ and $x($H$^{+}) \geq 30\%$).  The
high--ionization region is typical of density--bounded nebulae observed
with a large aperture.  The low--ionization and neutral regions are
characteristic of low--ionization features, such as ansae and FLIERs,
commonly found in planetary nebulae.  The computations stop at
$x($H$^{+}) = 30\%$ since the assumptions of ionization and thermal
equilibrium become poor at about that point and the use of {\sc Cloudy}
is not appropriate.  (Even at $x($H$^{+}) \approx 50\%$ the detailed
results are probably uncertain.)

The procedure was applied to a ``baseline model'' in which the total
density is 0 cm$^{-3}$ within a radius $R_{\rm o} = 10^{17}$ cm, and
3000 cm$^{-3}$ beyond it.  Abundances $\log
A($H$:$He$:$C$:$N$:$O$:$Ne$:$S$:$Ar$) =
0.0:-1.0:-3.52:-4.0:-3.22:-3.82:-4.82:-5.0$ were used.  A star with
$\log L_* = 37.4$ (erg s$^{-1}$)and a blackbody spectrum with
temperature $\log T_* = 5$ ($^{\circ}$K) was adopted.  This baseline
model is representative of typical bright galactic planetary nebulae
such as NGC 3242 and 7662.

Several input parameters were then varied in order to ascertain how the
derived abundances $A'(X)$ and $i_{CF}$ discrepancies ${\cal A}(X)$ are
affected.  The parameter variation studies are intended to explore
reasonable ranges of structural and chemical conditions.  These
parameter variation sets, $\Delta$, are summarized in the left column of
Table~\ref{tbl-1}.

The $\Delta R_{\rm o}$ sequence probes the effect of nebular geometry
through changes in the size of the empty central cavity.  For $\Delta$O,
$\Delta$N, and $\Delta$(N+O), abundances were varied to determine the
extent to which nebular abundances affect the $i_{CF}$ abundance
discrepancies.  We computed ten models, $\Delta (\log T_*,\log L_*$), in
which a ``star'' moves along a typical Harman--Seaton sequence in its PN
evolution.  That is, a range of stellar UV flux distributions (assumed
to be blackbodies) and luminosities was considered.  For comparison, we
also considered seven nonLTE models, $\Delta (\log T_*, \log g_*$),
which were chosen to approximate models from the $\Delta (\log T_*, \log
L_*$) sequence to examine the effects of changes in the shape of the
input spectrum.  These nonLTE models are interpolations of a set of
models from Klaus Werner (Werner {\it et~al}. 1991) included in the {\sc
  Cloudy} package.

\section{Results}

The model--computed ionization structure and emission line profiles of
the baseline model are shown in the top and center panels of
Figure~\ref{fig1}, respectively.  The figure is characteristic of nearly
all of the model nebulae.  The resulting abundance discrepancies, ${\cal
  A}(X,r)$, are shown in the lower panel of Fig.~\ref{fig1}.  Shaded
regions show the areas integrated for the low--ionization and neutral
regions.  Note that the differences between the ionization volumes for
O$^{++}$ and Ne$^{++}$ and O$^+$ and N$^+$ are cross--hatched in the
left and right panels, respectively.

The resultant radial profiles of $n'({\rm S}^+,r)$, $T'({\rm N}^+,r)$,
$T'({\rm O}^{++},r)$, and the ionic abundances $A'(X_{ion},r)$ are found
to be in such close agreement with their ``real'' counterparts computed
by {\sc Cloudy} that we need not distinguish between them hereafter.
This agreement is in spite of different atomic constants in the codes
used in both methods which only serves to reassure us that the errors in
the $i_{CF}$ method are real and measurable.  One exception is the
Ar$^{++}$ abundance determined from the 7136\AA \ line for which {\sc
  Ionic}'s results consistently differ from {\sc Cloudy}'s computed
values by a factor of two.  Other exceptions occur, but only in extreme
circumstances.

\subsection{Global ${\bf i_{CF}}$ abundances measurements.}

The emission line profiles were integrated over each of the four types
of regions (entire nebula, high--ionization, low--ionization, and
neutral) using both volume and line--of--sight weighting.  These results
mimic observations using apertures that isolate and average over the
corresponding volumes or lines of sight.  From these integrated
observables $n'({\rm S}^+)$, $T'({\rm N}^+)$, $T'({\rm O}^{++})$,
$A'(X_{ion})$, the $i_{CF}$ correction factors, the derived abundances
$A'(X)$, and $i_{CF}$ abundance discrepancies ${\cal A}(X)$ were derived
for each zone separately.

A grid of the line flux ratios integrated over the full nebular volume
and corresponding to each of the present model sequences is presented in
Table~\ref{tbl-1}.  The nebular H$\beta$ luminosity and radius for each
model (relative to the baseline) are tabulated, as are the densities and
temperatures that would be derived directly from full--volume integrated
emission line ratios.

It is interesting to note how the brightnesses of all of the
low--ionization lines increase (relative to H$\beta$) along the
Harman--Seaton sequence through all ten of the $\Delta (\log T_*, \log
L_*$) models.  By the end of this sequence, the [N{\sc ii}]/H$\alpha$
line ratio exceeds 1.5 (assuming a standard Balmer decrement value of
2.8 for the H$\alpha$/H$\beta$ line ratio).  Similarly, [O{\sc
  i}]/H$\alpha$ approaches 25\%.  These are global integrated values,
not just local ones near the nebular edge.  Comparing these resulting
flux ratios to shock models, we note that pure photoionization models
can give rise to low--ionization line strengths comparable to many
planar shock models with shock velocities $\leq 200$ km s$^{-1}$ (e.g.\ 
Hartigan {\it et~al}.\ 1994).  Within the potential uncertainties of
abundances determined with $i_{CF}$ methods, the line ratios in the
low--ionization portion of a planetary nebula can not reliably be
distinguished from shock models without additional information.

The H$\beta$ luminosities are relevant for studies of the PN luminosity
function.  In the case that the nebulae are ionization bounded, these
luminosities are insensitive to the assumed nebular geometry or
abundances (see Table~\ref{tbl-1}).  However, as expected, they are
excellent discriminators of the ultraviolet photon flux of the central
star, i.e.\ its evolutionary state.

The nebular densities and temperatures in Table~\ref{tbl-1} exhibit some
curious features.  The [S{\sc ii}] lines are weighted in favor of
regions that are typically only partially ionized.  Consequently $n({\rm
  S}^+)$ underestimates the true nebular electron density (3300
cm$^{-3}$ in this case) by more than 20\%.  In addition, $T({\rm N}^+)$
can differ from $T({\rm O}^{++})$ by more than 1000 $^{\rm o}$K
($\sim10\%$) due to changes in the nebular temperature profile through
the N$^{+}$ and O$^{++}$ regions.  Accordingly, observed $T({\rm N}^+)$
and $T({\rm O}^{++})$ can be expected to differ in real nebulae in which
no small--scale temperature fluctuations occur (cf Fig.~\ref{fig1}).
The sign of the $T$(N$^{+}$) -- $T$(O$^{++}$) temperature difference
reverses at stellar temperatures around 10$^5$ $^{\rm o}$K.

On the one hand, abundance errors are exacerbated unless $T({\rm N}^+)$
and $T({\rm O}^{++})$ are used for the computation of low-- and
high--ionization species, respectively.  On the other hand, the errors
which derive from problems in measuring $n'({\rm S}^+)$, $T'({\rm N}^+)$
and $T'({\rm O}^{++})$ are often small compared to other potential
problems which are inherent to the $i_{CF}$ method.

A key assumption in the $i_{CF}$ method is that the ionization structure
of N and O (and hence N$^{+}$ and O$^{+}$) is identical.  However, as
seen in Fig.~\ref{fig1}, {\it these ionization zones are obviously not
  cospatial!} Although charge transfer locks O$^{+}$ to H$^{+}$, charge
transfer is far less important in the ionization structure of N$^{+}$
(Osterbrock 1989).  Hence the ionization structures of N$^{+}$ and
O$^{+}$ are only weakly coupled.  Similarly, the ionization fractions
He$^{+}$, O$^{++}$, and Ne$^{++}$ are assumed to be roughly equal in the
$i_{CF}$ method.  In the {\sc Cloudy} models these deviate from one
another, especially near the edge of the high--ionization region.  Here
too, charge transfer affects O$^{++}$ to a greater degree than
Ne$^{++}$.  These potential errors in the $i_{CF}$ method have long been
recognized (e.g. P\'equignot {\it et~al}.\ 1978).

The substantive issue in this paper is a quantitative analysis of their
effect on $i_{CF}$ abundances, i.e.  the magnitude of the discrepancy,
${\cal A}(X)$.  The results of ${\cal A}(X)$ are summarized in
Table~\ref{tbl-2} for the $\Delta R_{\rm o}$ and $\Delta(\log T_*,\log L_*$)
sequences of models.  Results for the $\Delta {\rm O}$, $\Delta {\rm
  N}$, $\Delta ({\rm N}+{\rm O})$ sequences show only small departures
from those of the baseline model and are not shown.  Similarly, the
nonLTE model atmospheres don't change the resulting $i_{CF}$ abundances
to any great degree.  In general, the new discrepancies differ from the
blackbody case by $\leq 5\%$ for unresolved nebulae and $\leq 7\%$ for
resolved observations.  In cases where the blackbody abundances were
good (discrepancies $\leq 5\%$) the nonLTE result tends to differ by
$\leq 2\%$.

Several interesting conclusions can be drawn from inspection of the
table.  The abundances derived for He, N, and O from entire--volume
observations (i.e., large--aperture observations of ionization bounded
nebulae) are generally good.  A similar result was found by Henry {\it
  et~al}.\ (1996).  S and Ar are generally not fitted well (errors
$>20\%$) with Ar systematically overestimated by $>20\%$ in volume
integrals.  Errors for Ne can exceed $50\%$ or more and vary greatly
with the properties of the central star.

But the apparent accuracy of the $i_{CF}$ method can be deceptive.  Note
in Fig.~\ref{fig1} the degree to which ${\cal A}(X,r)$ varies with
radius.  In a volume--average over the entire nebula, these variations
tend to cancel; however, many of the local values of ${\cal A}(X,r)$
deviate from unity.  In regions of low ionizations the $i_{CF}$
discrepancies of He, N, Ne, and Ar often approach 20\%, and sometimes
even exceed 100\%.  Local $i_{CF}$ abundance discrepancies are
considered in \S\S3.2 and 3.3.

\subsection{Radial ${\bf i_{CF}}$ abundance measurements in the baseline model}

We next consider the abundances derived from observations in which the
baseline nebula is spatially resolved.  Refer to Fig.~\ref{fig1} and the
baseline model results shown in Table~\ref{tbl-2}. In the outer parts of
the nebula where the $i_{CF}$ approximation N/O = N$^{+}$/O$^{+}$ breaks
down, ${\cal A}($N$)$ increases from 0.7 in the nebular interior to 2 in
the neutral zone.  If the emission lines of the baseline model were
observed using a long slit, {\it the use of the $i_{CF}$ approximation
  would lead to the erroneous conclusion that both N/O and N/H increase
  by a factor of three or more!}

The abundances determined solely or largely from lines of high
ionization, He and Ne, also show large discrepancies in low--ionization
regions; however, unlike [N{\sc ii}], only a small fraction of the
helium and neon line fluxes arise outside the high--ionization zone.
Therefore the magnitude of the abundance discrepancy will not be
noticeable in practice.  Nonetheless, $i_{CF}$ abundance errors in He
and Ne will certainly arise in isolated low--ionization ansae, lobes, or
halos.

Similarly, the derived abundances of S and Ar using only the standard
[S{\sc ii}] and [Ar{\sc iii}] lines are affected. However, their
$i_{CF}$ discrepancies can be ameliorated if lines of [S{\sc iii}] and
[Ar{\sc iv}] are detected and included in the $i_{CF}$ abundance
analysis.

Clearly, except for O, all $i_{CF}$ abundances are vulnerable to large
systematic errors with nebular radius, especially in or near the
ionization front.  This is particularly apparent in the lower panel of
Fig.~\ref{fig1}.

\subsection{ Radial ${\bf i_{CF}}$ abundance measurements for other models}

The $i_{CF}$ discrepancies for other models show some interesting trends
from region to region.  Here we point out only a few of them.  Consider
first the $\Delta R_{\rm o}$ sequence.  As the radius of the central
hole grows and the impinging stellar radiation becomes more dilute
(i.e.\ the ionization parameter decreases), the relative size of the
low--ionization zone grows and the boundary between the high-- and
low--ionization regions becomes less distinct.  This has a profound
effect on abundances measured in the high--ionization zone, as seen in
Table~\ref{tbl-2}.  Hollow PNe such as Abell 39 and halos such as NGC
6826, ionized by highly dilute stellar radiation, are likely to be the
most affected.

Consider the $\Delta(\log T_*,\log L_*$) sequence.  Abundance
discrepancies are very sensitive to the assumed temperature of the
excitation source, as is easily seen in Table~\ref{tbl-2}.  The first
model in the sequence represents a cool stellar ionizing source, $\log
T_* = 4.55$.  In this case the high--ionization zone occupies less than
half of the nebular volume, and the boundary between it and the
low--ionization zone is not sharply defined.  In this transition region
the various ${\cal A}(X)$'s are large.  Only oxygen, for which both
important ionization stages are visible, has a reliably measurable
abundance.  This serves as a warning for abundance measurements of H{\sc
  ii} regions as well as low--ionization PNe such as IC418 and
BD$+30^{\rm o}3639$.

As the stellar temperature increases in the sequence the ionization
patterns seen in Fig.~\ref{fig1} are quickly established.  The
low--ionization zone becomes smaller in radius, and the lines that are
emitted from it drop in overall importance.  As a result, the abundance
discrepancies ${\cal A}(X)$ are closest to unity for the second through
fifth models in the sequence.

As the stellar luminosity drops (seventh through tenth models), the
low--ionization zone again fills more of the nebular volume.
Correspondingly, the ${\cal A}(X)$'s increase.  As discussed earlier,
the emission line ratios of low--ionization species such as O$^{\rm o}$,
S$^+$, O$^+$, N$^+$, and Ar$^{++}$ resemble values found in
low--velocity shocks (Hartigan {\it et~al}.\ 1994).  At the same time,
high--ionization lines from He$^{+}$, He$^{++}$, O$^{++}$, and Ne$^{++}$
remain bright.  Except for oxygen, the ionization correction factors
become large and the corresponding abundance discrepancies mount.

\section{Discussion}

Ionization correction techniques have long been used in both galactic
and extra--galactic astronomy.  As early as 1985, Bob Rubin recognized
problems with derived nitrogen abundances.  Our work bears out his
assertions.  We confirm that as electron temperature drops the ratio
N$^{++}$/O$^{++}$ rises.

While the present models focus on PNe, the $i_{CF}$ abundance
discrepancies are important for observations of galactic and
extragalactic H{\sc ii} regions (Gonzalez--Delgado {\it et~al}.\ 1994,
Martin 1996, and Storchi--Bergmann {\it et~al}.\ 1996 to name only a few
recent examples).  Storchi--Bergmann {\it et~al}.\ (1996), in their
examination of the central regions of AGNs, assume that $i_{CF}$
assumptions can lead to an under--estimate in the nitrogen abundance of
20\%; however, our results suggest that the discrepancy could be
40--50\% based on the nebular temperatures and densities that they find.

Nitrogen in radiation--bounded objects like planetary nebulae can be
easily overestimated, particularly in lines of sight that largely or
exclusively sample the low--ionization portion of the nebula where the
N$^+$ profile extends beyond both those of H$^+$ and O$^+$.  A slit
which lies across an ionization--bounded nebula should clearly show a
rise in N relative to both O and H due to the form of the $i_{CF}$ for
nitrogen.  This effect certainly contributes to the results of Balick
{\it et~al}.\ (1994) and Guerrero {\it et~al}.\ (1995), although it is
unclear whether or not it can completely account for the reported
variation in N/O.  Similar, although less pronounced, problems with Ne
and S also exist.  Surprisingly, the recent results of Corradi {\it
  et~al}.\ (1997) show the expected radial increase in all species {\it
  except} nitrogen.  These radial abundance discrepancies will be
exaggerated for highly extended nebulae such as bipolars in which the
flux of radiation striking most of the lobe walls is both dilute and
incident at high obliquity (see the values for the $\Delta R_{\rm o}$
series in Table~\ref{tbl-2}).

Since 1967, Peimbert, Torres--Peimbert and their collaborators have been
examining the physical effects of both large-- and small--scale
temperature fluctuations on abundances determined by the $i_{CF}$
method.  Recently, Peimbert {\it et~al}.\ (1995b) used emission line
temperatures from different elements to attempt to measure the
temperature fluctuations in 17 Type I PNe.  To quantify the
fluctuations, they measured temperatures for numerous emission lines and
compared them to one another.  Our work clearly shows that this is not a
sound test as the lines originate at different points in the nebula due
to different sensitivity to temperatures and densities.  An examination
of the temperatures in Table~3 of Peimbert {\it et~al}.\ (1995b) shows
that roughly half of the objects (notably including all of the Minkowski
objects in their sample) have $T({\rm N}^+)-T({\rm O}^{++})$ values that
are within the differences that one should expect to measure in a
smooth, homogeneous PN with no small--scale temperature fluctuations.

Recently, a puzzling factor--of--three--to--five disagreement in the
oxygen abundances measured from forbidden and recombination lines has
arisen (Liu \& Danziger 1993, Peimbert {\it et~al}.\ 1993, Liu {\it et
  al}.\ 1995, Mathis 1995).  The present study shows that O abundances
measured using the $i_{CF}$ method are generally accurate to $\sim5\%$
measured globally, and 10\% measured with a pencil beam near the nebular
center (before observational errors).  Thus our results do not resolve
this controversy.

\section{Conclusions}

The present model results show that use of $i_{CF}$ techniques for abundance
analysis for spatially unresolved objects gives results accurate to
$\sim20\%$ as found by Henry {\it et~al}.\ (1996).  Larger discrepancies
can occur for some nebular conditions including cool ionization sources
or highly dilute incident stellar UV radiation.

For spatially resolved nebulae, $i_{CF}$ abundance errors can be high
locally, especially for N, Ne, and Ar in or near an ionization front.
These errors have led to probable overestimates of N/O abundance
variations in FLIERs and bipolar PNe.

Further, the emission line ratios predicted for the low--ionization
species in or near ionization fronts in resolved observations resemble
those predicted for shocks.  This is the cause of some ambiguity, if not
confusion, in the interpretation of line ratios.

The magnitudes of errors in the $i_{CF}$ method differ for nebulae of
various sizes and excitation/ionization sources.  In addition, the
$i_{CF}$ errors depend on the extent to which ionizing photons are
absorbed or escape (i.e.\ whether the nebula is ionization or density
bounded).  Consequently, statistical studies of abundances in many
objects need to account for these types of conditions.

Nebulae excited by stars with $T_* \leq 45000^{\rm o}{\rm K}$ have
relatively large $i_{CF}$ discrepancies.  Our work has concentrated on
conditions typical in PNe and thus suggests that a detailed study of
$i_{CF}$ abundances in galactic and extragalactic H{\sc ii} regions
needs to be undertaken in order to assess the veracity of
$i_{CF}$--derived abundances and gradients in the disks of spiral
galaxies.

It is a pleasure to thank the developer of {\sc Cloudy}, Gary Ferland.
This research was sponsored by NSF grant AST 9417112 and NASA GO-6117.

\newpage
\appendix
\section{Empirical Procedures for Deriving Abundances}

\subsection{Densities and Temperatures}

Straightforward analytical expressions can be used to derive densities,
temperatures and ionic abundances from the measured emission line
ratios.  The functions shown below were obtained by a linear
least--squares fit to a series of results from the {\sc Ionic} and {\sc
  Temden} tasks in the IRAF/STSDAS software package.  I$\lambda$ or
I($\lambda\lambda$) is the measured intensity of a line relative to
H$\beta = 1$ (not 100) after correction for reddening.

The task {\sc temden} in the nebular package in IRAF returns a value of
density or temperature for a specific set of diagnostic line ratios.
The density formula is computed assuming $T = 10^4$ $^{\rm o}$K and is
valid for $100<n({\rm cm}^{-3})<30000$.  Assuming that the atomic data
are correct, the errors of the fits for $n({\rm S}^+)$ and $n({\rm
  Cl}^{++})$ are typically 5\% or less.  Those for $n({\rm Ar}^{+++})$
are about 8\%.  As in the paper above, X$^+$ implies X$^+$/H$^+$ and X
implies X/H.

\noindent $\displaystyle \log[n({\rm S}^+)] = 21.683 - 41.177r
+ 39.848r^{2} - 13.954r^{3}$ \hfill where $\displaystyle
r=\frac{I6717}{I6731}$

\noindent $\displaystyle \log[n({\rm Cl}^{++})] = 14.775 - 19.036r
+ 20.275r^{2} - 8.3713r^{3}$ \hfill where $\displaystyle
r=\frac{I5517}{I5538}$

\noindent $\displaystyle \log[n({\rm Ar}^{+++})] = 12.713 - 5.7379r
- 5.9903r^{2} + 14.375r^{3} - 7.1339r^{4}$ \hfill where $\displaystyle
r=\frac{I4711}{I4740}$

\noindent The temperature formulae are computed assuming
$n=10^3{\rm cm}^{-3}$ and are valid for $6000<T(^{\rm o}{\rm K})<17000$.
Errors are less than 1\%.

\noindent $\displaystyle \log[T({\rm N}^+)] = 5.96r^{-0.387} +
8.13$ \hfill where $\displaystyle r=\frac{I(6548+6583)}{I5755}$

\noindent $\displaystyle \log[T({\rm O}^{++})] = 5.22r^{-0.304} + 8.185$
\hfill where $\displaystyle r=\frac{I(4959+5007)}{I4363}$

\noindent Normally some quality--weighted average value of $n$ and $T$
are adopted before proceeding further.  However, if the data are good,
then values of $n$ and $T$ can be used which are optimized for the
ionization potentials of the ions in the expressions below.

\subsection{Ionic Abundances}
Ionic abundances of He are computed from expressions fitted to effective
recombination coefficients found in Osterbrock (1989) (Tables 4.4, 4.5,
and 4.6).  A density of $n=10^{4}$ ${\rm cm}^{-3}$ has been assumed.  We
acknowledge that recent work (e.g.\ Smits 1996) has provided better
coefficients for helium, however since our goal is to identify errors in
the method, rather than differences in atomic coefficients, we have
chosen to use Osterbrook's values which match those used in {\sc
  CLOUDY}.  The deviation from curves fit to Osterbrock's tables is
smaller than 5\%.

The task {\sc ionic} in the nebular package in IRAF was used to compute
various ionic abundances from forbidden line ratios (relative to
H$\beta$) at $T=7500$, $10000$, $12500$, and $15000$ $^{\rm o}$K for
each of the following densities: $30$, $100$, $300$, $1000$, $3000$,
$5000$, $10000$, $15000$, $20000$, $25000$, and $30000$ ${\rm cm}^{-3}$.
Empirical expressions for N, O, Ne, S, and Ar isotopes whose
coefficients (including those inside the exponentials) are determined
from a least--squares fit to IONIC's results as well as those for He
appear in Table~\ref{tbl-3}.  These fits are not accurate outside of the
ranges of temperature and density above.

In Table~\ref{tbl-3}, $x=n/\sqrt{T}$ and line ratios are all relative
to H$\beta=1$, not 100.  To the right side of the fits, residuals of the
fit are shown at $7500$, $10000$, $12500$, and $15000$ $^{\rm o}$K,
respectively, as a percentage of the fitted value.

\subsection{Total abundances}

The ionic abundances are multiplied by ionization correction factors to
derive total abundances.  The primary source for the expressions in
Table~\ref{tbl-4} is Kingsburgh \& Barlow (1994; KB).  Alternate sources
of ionization correction factors include K\"{o}ppen {\it et~al}.\ (1991;
KAS), Peimbert {\it et~al}.\ (1995b; PLTP) and Freitas--Pacheco {\it et
  al}.\ (1995; F--PBCI).

\clearpage 

\figcaption[fig1.ps]{The ionization structure (upper panels; from {\sc
    Cloudy} 90.02), line emissivities (middle panels) and
  $i_{CF}$--derived abundance discrepancy (lower panel) for the baseline
  model.  The shaded regions are zones where O$^{\rm o}$, O$^+$,
  O$^{++}$ and higher ionization species dominate, respectively.  The
  inner edge of the nebula is at $10^{17}$ cm. \newline \hspace*{.75 in}
  Cross hatching in the upper left panel shows deviations from the
  $i_{CF}$ assumption that O$^{++}$ and Ne$^{++}$ are cospatial.
  Similarly, cross hatching in the upper right panel indicates
  deviations from the assumption that N$^+$ and O$^+$ are cospatial.  In
  the lower panel, notice the tendency for the discrepancies to cancel
  when integrated over large radii or volumes.  This implies that
  $i_{CF}$ abundances are more likely to be specious when sampled
  through smaller apertures.
  \label{fig1}}

\begin{table}
\dummytable\label{tbl-1}
\end{table}
 
\begin{table}
\dummytable\label{tbl-2}
\end{table}
 
\begin{table}
\dummytable\label{tbl-3}
\end{table}

\begin{table}
\dummytable\label{tbl-4}
\end{table}

\end{document}